# Quantum Emergence of Linear Particle Accelerator and Anomalous Photon-induced Near-field Electron Microscopy in a Strong Coupling Regime

Jing Zhou[1], Ido Kaminer[2], Yiming Pan[3]

1. Department of Science, Chongqing University of Posts and Telecommunications, Chongqing 400065, China
2. Department of Electrical Engineering, Technion, Israel Institute of Technology, Haifa 3200003, Israel
3. Department of Physics of Complex Systems, Weizmann Institute of Science, Rehovot 76100, ISRAEL

Correspondence and requests for materials should be addressed to Y.P. (yiming.pan@weizmann.ac.il).

**Abstract**

Photon-induced near-field electron microscopy (PINEM) is a currently developing spectral approach that characterizes quantum electron-light interactions in electron energy gain/loss spectrums, with symmetrically discretized gains or losses of light quanta ($\hbar\omega$), coupled with a laser induced optical near-field. In this letter, we have demonstrated that Linear Particle Accelerator (LPA) and anomalous PINEM (APINEM) can analytically emerge from PINEM-kind interaction in a strong coupling regime, because of quantum interferences of photon sidebands overlap. Furthermore, we also found that the pre-interaction drift (or free propagation) in point-particle regime can produce interesting optical spectral focusing and periodically spectral bunching of electron energy/momentum distribution, which enable us to improve the spectral resolution of electron imaging and spectroscopy. These observation of LPA and APINEM in strong laser physics can be of great interests for both theoretical and experimental communities, such as ultrafast electron microscopes, attosecond science and laser-driven accelerators.

Photon-induced near-field electron microscopy (PINEM) is strong field light-matter interaction physics that can coherently manipulate with light, quantum electron wavefunctions in sub-femtosecond or attosecond scales [1]. It involves the discretized multiphoton emission and absorption processes that take place when a single free electron pulse passes through the optical near-field of a nanostructure or metamaterial such as a gold needle tip that is illuminated by a femtosecond IR laser pulse (800 nm) [2]. The schematic setup of a tip-like PINEM interaction is shown in Fig.1a. The single electron pulse is ejected from a photoemission electron gun and freely drifts with length $L_0$ to arrive at the entrance of a small near-field interaction region [3-5]. The optically excited tip-like interaction length L is in the scale of several tens of nanometers. In the electron-photon interaction region, the quantum electron dramatically exchanges photons of energy quanta $\hbar\omega$ with the optical near-field, and then acquires spectral or temporal modulations when leaving the interaction region and drifting with distance $L_D$ until being observed in an electron energy loss spectrum (EELS). Following the experimental setup of C. Ropers et al. [2,6], we modeled the quantum electron multi-photon interaction in a strong coupling regime by solving the relativistically modified Schrodinger equation with the near-field perturbation Hamiltonian [7]

$$H = H_0 - \frac{e}{m} p \cdot A = H_0 + \frac{eF(z)}{\gamma m \omega} p \sin(\omega z/v_0 - \phi_0) \qquad (1)$$

where $H_0 = E_0 + v_0(p - p_0) + (p - p_0)^2/2m^*$, $m^* = \gamma^3 m$ is the effective electron mass in longitudinal direction (z) and the initial electron energy $E_0 = c\sqrt{m^2c^2 + p_0^2} = \gamma mc^2$ and momentum $p_0 = \gamma m v_0$ with the Lorentz factor $\gamma = \left(\sqrt{1-\beta^2}\right)^{-1} = 1.4$ and the velocity ratio $\beta = v_0/c = 0.7$, $\omega$ is the optical frequency of the laser pulse, and F(z) is the slow-varying spatial field distribution ($\sim 10^{6-8}$ V m$^{-1}$) of the laser-illuminated optical excitation. Note that we ignored the contributions of the electron's transverse components. The relative phase ($\phi_0$) between the optical near-field and the electron pulse is controlled by the phase locking between the two laser pumps at the photocathode and at the metallic tip [2,8-10]. With the approximation of the strong coupling regime, one can obtain the modulated momentum (or energy) distribution of the PINEM-kind interaction

$$\psi_p = (2\pi\sigma_p^2)^{-1/4} \sum_n J_n(2|g|) e^{-in\phi_0} \exp\left(-\frac{(p-p_0-n\delta p)^2}{4\sigma_p^2}\right), \qquad (2)$$

where $\delta p = \hbar\omega/v_0$ is the discretized correspondent to the light quanta $\hbar\omega = 1.55\ eV$ ($\lambda = 800$ nm). We also take energy conservation of photon exchanging ($E_f - E_0 - n\hbar\omega = 0$) in multiphoton absorption and emission where n is an integer and $\sigma_p = \sigma_E/v_0$ is the intrinsic electron wavepacket momentum width where the energy uncertainty $\sigma_E = 0.3\ eV$ (i.e., FWHM is $0.7\ eV$) in the UTEM experiment [2,10]. From the Heisenberg uncertainty relation, we can estimate the intrinsic spatial wavepacket size of free electron is $\sigma_{z_0} = \hbar/2\sigma_p$. The Bessel function $J_n$ relates to the multiphoton process of the $n^{th}$-order scattering, and the normalization of the modulated electron wavefunction is automatically satisfied with its mathematical property $\sum_n |J_n(2|g|)|^2 = 1$ for any real argument $|g|$. Finally, the short-time approximation is taken into account for the tip-like near field interaction ($< 10^{-15}$ s) and the participated photon number is effectively given by $2|g| = \int_0^L \frac{eF(z)}{\hbar\omega} dz$.

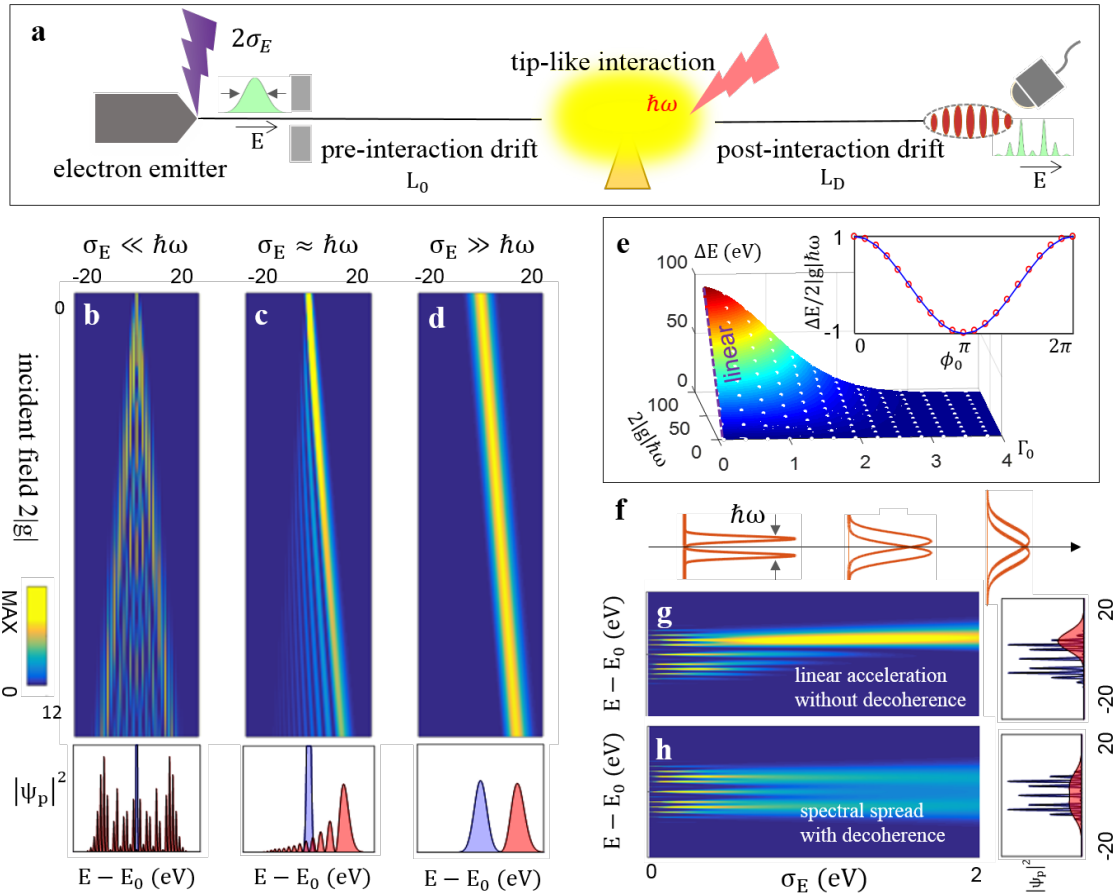

Figure 1| (a) The strong-field light-electron interaction setup for photon-induced near-field electron microscopy (PINEM) and linear particle accelerators (LPA). (b-d) The transition from PINEM to LPA regimes as a function of incident field strength $2|g|$ at conditions $\sigma_E \ll \hbar\omega, \sigma_E \approx \hbar\omega, \sigma_E \gg \hbar\omega$, respectively, where $\hbar\omega = 1.55\ eV, \phi_0 = 0$. Correspondingly, the typical spectral distributions are plotted below at $2|g| = 12$. Quantum emergence of the laser-driven accelerator is achieved at the point-particle limit $\sigma_E \gg \hbar\omega$. (e) The wavepacket acceleration as a function of the incident field energy $2|g|\hbar\omega$ and the decay parameter $\Gamma_0 = \hbar\omega/2\sigma_E$. The inset shows the acceleration dependence of the relative phase ($\cos\phi_0$). (f) The electron wavepacket acceleration originates from the overlap between the spectral photon

sidebands. (g-h) demonstrates the PINEM-to-LPA transition as a function of energy uncertainty ($\sigma_E$) without decoherence, and with decoherence, respectively, where (g) shows the linear acceleration but (h) spectrally broadens incoherently with only a partial acceleration region.

**Quantum emergence of linear particle accelerator (LPA)** – The first surprising result, as we presented in this Letter, is that the classical acceleration emerges from the quantum interference between PINEM sidebands in a strong coupling regime. If one allows to alter the ratio between the intrinsic momentum uncertainty ($\sigma_p$, or energy width $\sigma_E$) and the discretized momentum spacing ($\delta p$, or light quanta $\hbar\omega$ in EELS), the electron energy loss spectrum (EELS) observation of the modulated momentum/energy density distribution $|\psi_p|^2$ (Eq. 2) would be dramatically influenced. Fig.1b-d show the normalized spectral distributions as a function of the incident field strength ($2|g|$) for three cases $\sigma_p \ll \delta p, \sigma_p \approx \delta p$ and $\sigma_p \gg \delta p$, respectively. Fig. 1b shows the typical PINEM with obvious symmetric photon sidebands, which has been nicely observed by Feist et al. [2,10]. In their experiment, the energy width of the single electron pulse is filtered to $\sigma_E = 0.3\ eV$ that is smaller than the photon quanta $\hbar\omega = 1.55\ eV$, in which case that the interacting electron with light is in the plane-wave picture. However, if we coherently vary the intrinsic energy width close to or even more than $\hbar\omega$, the EELS observations in the Fig. 1c and 1d become unexpectedly non-symmetric. We find that the net electron acceleration is linearly proportional to the incident field strength ($2|g|$) at the condition $\sigma_E \gg \hbar\omega$ as shown in Fig. 1e. This condition corresponds to the point-particle interaction picture for a quantum free electron [7,11]. In addition, Fig. 1b-d demonstrates the possible intermediate transition characteristics from PINEM to LPA, especially, in which an Airy-like spectral pattern is found at the condition $\sigma_p \approx \delta p$. Note that these typical Wigner functions in the transition process from PINEM to LPA can be found in Fig. S1 of the SM file.

Thus, we expect the energy (or momentum) transfer of quantum wavepacket acceleration to follow

$$\Delta E = \int |\psi_p|^2 (E_p - E_0)\, dp = 2|g|\hbar\omega \cos(\phi_0)\, e^{-\Gamma_0^2/2}, \qquad (3)$$

where the decay parameter $\Gamma_0 = \delta p/2\sigma_p = \hbar\omega/2\sigma_E = \omega\sigma_{t_0}$ and $\sigma_{t_0} = \sigma_{z_0}/v_0$ is the intrinsic electron wavepacket duration. The net energy transfer ($\Delta E$) after the strong field interaction is then calculated at $2|g| \gg 1$, as shown in Fig. 1e. The extinction coefficient ($e^{-\Gamma_0^2/2}$) as a function of the decay parameter ($\Gamma_0$) demonstrates the intrinsic momentum uncertainty ($\sigma_p$) of a single electron which has a physical effect in its interaction with the strong laser field and matter [7]. The direct calculation (points) from Eq. (2) matches perfectly with our analytical acceleration formula (surface plot) from Eq. (3). The maximal energy gain is achieved in the point-particle picture of the electron ($\omega\sigma_{t_0} \ll 1$) that linearly equals to $2|g|\hbar\omega = \int_0^L eF(z)\,dz$, which is effectively associated with the classical acceleration gradient from the interacting near-field [8]. As a result, the central-shifted momentum distribution (Fig. 1d) in point-particle condition $\Gamma_0 \ll 1$ is then given by

$$\left|\psi_p^{(LPA)}\right|^2 = \left(2\pi\sigma_p^2\right)^{-1/2} \exp\left(-\frac{(p-p_0-2|g|\delta p)^2}{2\sigma_p^2}\right). \qquad (4)$$

Indeed, this is the laser-induced quantum emergence of the classical linear particle acceleration (LPA) in a strong coupling regime. This result can also be directly induced from classical electrodynamics as the moving free electron that is assuming a charged particle in the presence of an electromagnetic field. Here, we observe that the 'classical' point-particle acceleration behavior of a quantum wavefunction emerged as a quantum phenomenon, even in multiphoton interaction processes ($2|g| \gg 1$). Conversely, in the plane-wave operating condition $\Gamma_0 \gg 1$, there is no net gain in EELS observation (i.e., $\Delta E = 0$) but the symmetric PINEM sideband distribution $\left|\psi_p^{(PINEM)}\right|^2 = \left(2\pi\sigma_p^2\right)^{-1/2} \sum_n |J_n(2|g|)|^2 \exp\left(-\frac{(p-p_0-n\delta p)^2}{2\sigma_p^2}\right)$, in which approximately, the Gaussians can be reduced to delta functions $\delta(p - p_0 - n\delta p)$ with the n$^{th}$ order photon sideband yielding the measured probability $P_n = |J_n(2|g|)|^2$ [2]. Since the PINEM photon sidebands are almost symmetric (ignoring the recoil effect), the very existence of the net energy transfer (Fig.1e) is hard to believe. In the point-particle limit, however, those photon sidebands begin to overlap, and then the quantum interference between different sidebands reshape the final spectral distribution and produce the net electron acceleration [7].

To explain the quantum-to-classical transition from the symmetric PINEM to the anti-symmetric LPA, we have to consider quantum interference when the spectral sidebands overlap in the condition $\sigma_E > \hbar\omega$. Fig. 1f schematically shows the quantum overlap of two neighboring sidebands (with fixed spacing $\delta p$). When the two sidebands begin to overlap, such as the cross term of the Bessel functions $J_n J_{n+1}$ in $|\psi_p|^2$ (Eq. 2) will be relevant to the anti-symmetry of LPA, because of the relation $J_{-n}J_{-n-1} = (-1)^{2n+1}J_n J_{n+1} = -J_n J_{n+1}$ at $\phi_0 = 0$. We notice that all the interference terms are anti-symmetric for the spectral absorption-acceleration and emission-deceleration regions (i.e., n > 0 for absorption, n < 0 for emission), which eventually produce the net energy transfer. However, if $\phi_0 = \pm\pi/2$, we obtain no acceleration due to $J_{-n}J_{-n-1}e^{\mp i\pi/2} = J_n J_{n+1} e^{\pm i\pi/2}$, as well as from Eq. 3 when $\phi_0 = \pm\pi/2$. This phase dependence ($\phi_0$) is depicted in the inset of Fig.1e. Correspondingly, Fig. 1g shows the final distribution $|\psi_p|^2$ with the PINEM to LPA transition as increasing the intrinsic energy uncertainty $\sigma_E$ since the sidebands overlap becomes relevant to the EELS measurement [8,9]. We address that the quantum interference between the photon sidebands is probably sensitively comparable to the circumstances of ultrafast electrons (Fig.1a), along with the parameters of optical near-field configuration. The environment-induced decoherence can kill all those interferences during the multiphoton scattering and merely yields the mixed state of photon sidebands (or the reduced density matrix). As a result, with the energy uncertainty increasing, the final EELS distribution acquires a spectral spread but having no energy gain or loss, as shown in Fig.1h.

**Optical spectral focusing from anomalous photon-induced near-field electron microscopy (APINEM)** – Secondly, to achieve the exact anomalous PIENM [11], the pre-interaction free drift has to be taken into account. If we introduce a pre-interaction drift duration ($t_0$) into the initial electron wavepacket before near-field interaction, we expect

$$|\psi_p(t_0)|^2 = \left|(2\pi\sigma_p^2)^{-\frac{1}{4}} \sum_n J_n(2|g|) e^{-in\phi_0} \exp\left(-\frac{(p-p_0-n\delta p)^2}{4\sigma_p^2(1+i\xi t_0)^{-1}}\right)\right|^2, \quad (5)$$

where $\xi = 2\sigma_p^2/m^*\hbar$ is the chirp parameter and the per-interaction drift duration $t_0 = L_0/v_0$. Each Gaussian envelope in Eq. (5) is a pre-chirped wavepacket. In a strong coupling regime (i.e., $|g| = 3$), here, we demonstrated the laser-induced spectral focusing as a possible implement in APINEM regime. Fig. 2 shows the EELS observation as increasing the per-interaction drift length $L_0$. The initial spectral uncertainty ($\sigma_E = 7.8\ eV$) is extremely large as compared to the light quanta $E_p \gg \hbar\omega$. Fig. 2a-d demonstrate the focusing, accelerated focusing, defocusing, and decelerated focusing spectral patterns for the relative phases $\phi_0 = -\frac{\pi}{2}, 0\frac{\pi}{2}, \pi$, respectively. The final achieved energy uncertainty has shrunk to $\sigma_E^{(f)} = 2.6\ eV$ at the focusing position at $\phi_0 = -\pi/2$, in which the corresponding spectral resolution is enhanced approximately three times than that of the incident electrons. Fig. 2e shows the spectral focusing as a function of pre-interaction drift, in which the optimal spectral focusing emerges at the drift length $L_{0,opt} = 0.8\ cm$.

The optical spectral focusing of APINEM is honestly beyond our expectation. This anomalous effect, as well as the emergent LPA, comes from the quantum overlap between the chirped sidebands in the condition $\sigma_E \gg \hbar\omega$, and as a novel methodology, it enables us to improve energy resolution in electron microscopy and spectroscopy embedded on the coherently manipulation of the quantum interference of electron wavefunction in strong laser physics. More details of spectral focusing are in Fig. S2-3 of the SM file.

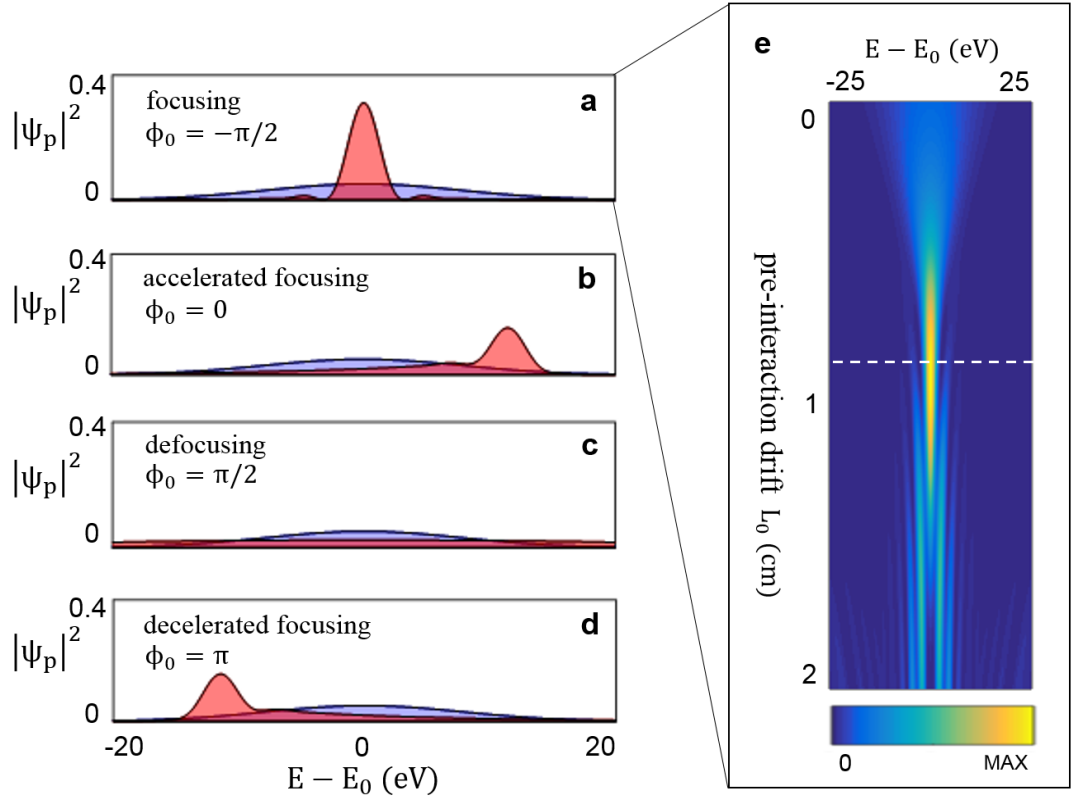

Figure 2| Achievement of the spectral focusing and high-resolution electron energy loss spectrum (EELS) in anomalous photon-induced near-field electron microscopy. (a-d) show the four typical cases: focusing, accelerated focusing, defocusing and decelerated focusing of initial relative phases $\phi_0 = -\frac{\pi}{2}, 0, \frac{\pi}{2}, \pi$, respectively, in optimal pre-interaction drift length (or duration) $L_{0,opt}$ and a strong field acceleration regime with parameters $\hbar\omega = 1.55\ eV, \sigma_E = 7.8\ eV, |g| = 3$. (e) demonstrates the spectral focusing as a function of pre-interaction drift $L_0$ at $\phi_0 = -\frac{\pi}{2}$.

**Momentum/energy periodically bunching vs. spatial/temporal periodically bunching -** The free drift of quantum free electrons before or after laser-induced interactions have different consequences. To address, we now compared the pre-interaction and post-interaction chirp effects in the point-particle interaction regime ($\Gamma_0 \ll 1$) and plane-wave interaction regime ($\Gamma_0 \gg 1$), respectively. Fig. 3a-c shows the typical periodically bunched energy/momentum distribution as a function of pre-interaction drift in APINEM regime. The relevant parameters are $|g| = 0.3, \Gamma_0 = 0.13$. As we known, the spectral sideband spacing in PINEM regime equals to light quanta ($\hbar\omega$, or $\delta p$, Eq. 2), but that in the APINEM regime it is completely different [11]. In fact, the spectral spacing of APINEM is proportional to the optical wavelength, meaning, inversely proportional to the light quanta. This anomalous spectral spacing can be geometrically explained in phase space representation as discussed in ref. [11]. The linear dependence of APINEM momentum spacing to the optical wavelength and its inversely dependence to the pre-interaction drift length $L_0$, is approximated to be $\delta p^{(APINEM)} = \beta\lambda(m^*v_0/L_0)$ [11], while the PINEM sideband spacing is given by $\delta p^{(PINEM)} = 2\pi\hbar/\beta\lambda$.

Conversely, the free-space evolution of the modulated electron wavefunction in spatial and temporal space with post-interaction drift duration $t_D$ is given by $\psi(z, t_D) = \frac{1}{\sqrt{2\pi\hbar}} \int \psi_p(0) e^{ipz/\hbar} e^{-iE_p t_D/\hbar} dp$, for which case the setup is shown in Fig.3d. We draw the periodically bunched density distribution $|\psi(\zeta, t_D)|^2$ in Fig. 2e as a function of the moving frame of reference $t - z/v_0$ and the post-interaction drift length $L_D = v_0 t_D$. The parameters of the setup are $|g| = 1, \sigma_z = 1.5\ \mu m, L_0 = 0$. Correspondingly, Fig. 3f shows the explicit under-bunching, optimal bunching and over-bunching of density distributions in sub-femtosecond or attosecond scales at post-interaction drift lengths $L_D = 1\ cm, 1.8\ cm$ and $4\ cm$, respectively.

In essential, it is interesting to compare the two different periodical bunching effects in the APINEM and PINEM regimes. Indeed, both the spectral and attosecond density bunchings originate from quantum interference fringes between photon sidebands as presented in phase space. In the setup with comparable pre-chirped propagations and weak near-field interactions (Fig.3a), the momentum projection of interference fringes into momentum/energy domain which leads to APINEM [11]. In the setup with PINEM interaction and post-chirped propagations (Fig.3d), while, these projection fringes instead emerge into spatial/temporal domain yielding the attosecond density bunching. Note that the post-interaction drift cannot shape the EELS observation because the momentum is a good quantum number in free space, instead, it can produce (periodically) density bunching [2]. Attosecond or sub-femtosecond density bunching have been widely observed by several experimental groups [10,14,15]. However, APINEM has not been reported or found experimentally, even though it is capable of substantially improving the high spectral resolution of electron microscopy and spectral periodic bunching.

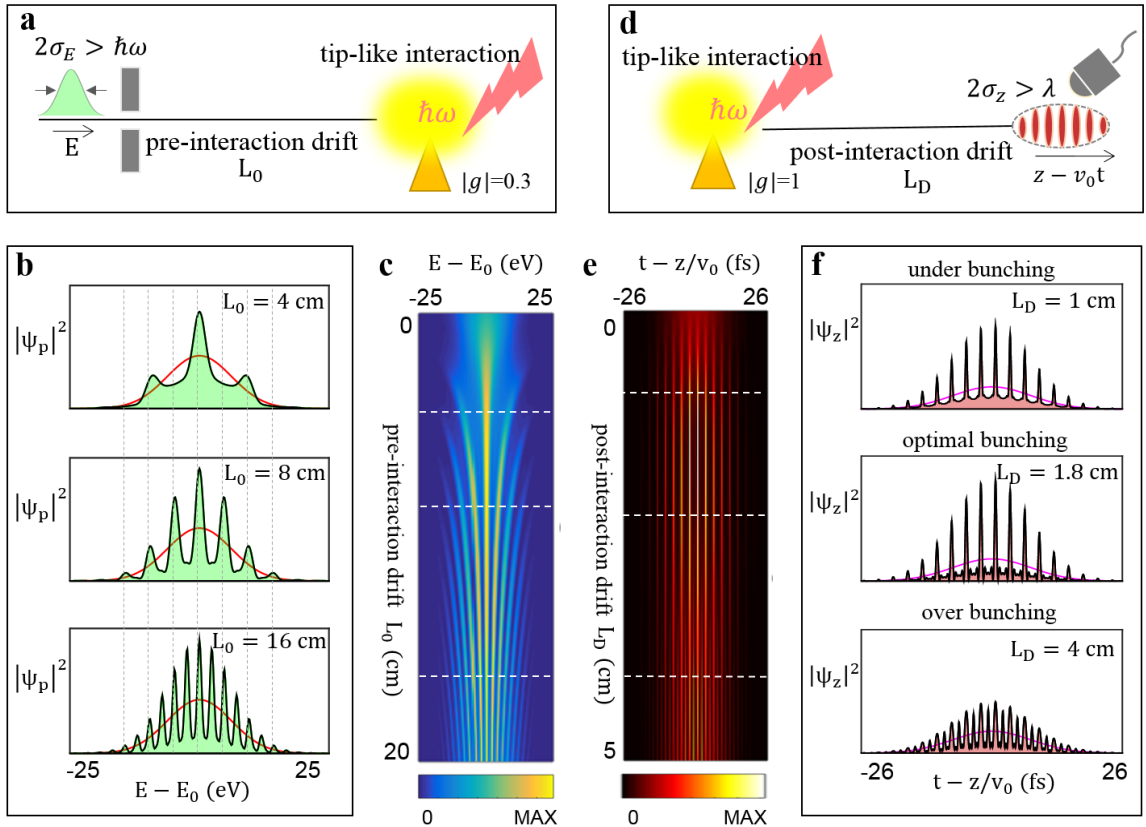

Figure 3| Comparisons between the momentum/energy bunching (a-c) and the spatial/temporal bunching (d-f). (a) The setup is composed of a pre-interaction drift and a tip-like near-field interaction, where the laser wavelength $\lambda = 800\ nm$ ($\hbar\omega = 1.55\ eV$) (b) The spectral periodic bunching in EELS appears as increasing for pre-interaction drift lengths $L_0 = 4\ cm, 8\ cm, 16\ cm$, respectively. The APINEM spacing is inversely proportional to $L_0$ in the centimeter scale. (c) shows the density plot of EELS as a function of pre-chirped length $L_0$. (d) the setup is composed of a tip-like interaction and a post-interaction drift; (e) shows the temporal density plot of time-of-flight as a function of post-chirped length $L_D$ in the distance of a centimeter scale. (f) The spatial/temporal periodic bunching in sub-femtosecond or attosecond scale at $|g| = 1.0$ is demonstrated in the cases of under-bunching ($L_D = 1\ cm$), optimal bunching ($L_D = 1.8\ cm$) and over-bunching ($L_D = 4\ cm$), respectively.

**Relationships between PINEM, LPA and APINEM regimes** – As a big picture, Fig. 4 depicts the transition regimes of these three typical spectra as a function of the interaction field wavelength and the per-chirped drift of electron at the uncertainty $\sigma_E = 0.3$ eV and the strong coupling $|g| = 10$. In PINEM regime, we observe a bunch of multi-photon sidebands with spectral portions given by $P_n = |J_n(2|g|)|^2$, as shown in Fig. 4a (also in Fig. 1b), in which the interaction field is in the optical infrared range ($\hbar\omega = 1.55\ eV, \lambda = 800\ nm$). At this moment, both the pre-interaction and the post-interaction drifts cannot shape or modulate the EELS spectrum, and therefore, Fig. 4c shows the pre-chirped PINEM spectrum with distance $L_0 = 12$ cm that is same as the unchirped case. Interestingly, for fixed electron intrinsic uncertainty ($\sigma_E = 0.3$ eV, i.e., FWHM is 0.7 $eV$), Fig. 4b shows the unchirped wavepacket acceleration as we expect in the point-particle condition $\Gamma_0 \ll 1$ for the interaction of optical filed in THz range ($\hbar\omega = 5\ meV, \lambda = 0.25\ mm$), and meanwhile Fig. 4d shows the APINEM spectral bunching at the acceleration regime but with pre-interaction drift ($L_0 = 0.23$ cm). It indicates that a same electron can act like a quantum wave-like interaction with the infrared field, but also interacts as a classical point-like particle in THz range. Besides, the spectral focusing or bunching of APINEM in THz range is extremely energy resolved in few meV scale, beyond the highest energy resolution of art-of-the-state electron microscopes [6]. As a result, we can conclude that the very existence of LPA and APINEM requires the intrinsic interaction condition $\sigma_E > \hbar\omega$.

Note, that in the weak field interaction case (i.e., $|g| \ll 1$), only the single-photon emission and absorption are relevantly involved, which means that the higher-order photon scatterings are negligible as $J_n(2|g|) = 0$ at $|n| \geq 2$. The final momentum components in Eq. (2) thus can be approximated as $\psi_p \approx (1 - |g|^2)\psi_p^{(0)} + |g|e^{-i\phi_o}\psi_{p-\delta p}^{(0)} - |g|e^{i\phi_o}\psi_{p-\delta p}^{(0)}$, where we take the Tayler series of Bessel functions $J_0(2|g|) \approx 1 - |g|^2, J_{\pm 1}(2|g|) \approx \pm|g|$. The weak field approximation would lead to the same results from the first-order perturbation analysis based on a grating structure [11]. We stress here that our extensive observation is general both for the weak field interaction ($|g| \ll 1$) and the extremely strong coupling regime ($|g| \gg 1$).

Freely changing the intrinsic electron uncertainty ($\sigma_E$, or $\sigma_z$) as compared with the light quanta ($\hbar\omega$ or $\lambda$) is essential to observe the PINEM to LPA transition experimentally. It should be noted, however, that limited studies are concentrated on investigating the intrinsic uncertainty of photoelectrons [16-18]. According to the work of P. Baum et al. [16], the intrinsic uncertainty of a single electron pulse relates to the tunable laser pulse parameters, accelerated field and electron gun (cathode) configuration when the free electron is being produced above the cathode's work function. Conversely, we can introduce the presently available tunable femtosecond laser pulse with varying light wavelength $\lambda$ to achieve the PINEM-to-LPA transition ($\Gamma_0$) that appears more reasonable and flexible. For instance, the THz technique is more flexibly applicable compared to the optical frequency range because the THz wavelength is much longer and easily controlled for the typical electron pulse configurations [15]. Besides, the spectral focusing and anomalous periodically bunching in APINEM regime come into observation when the pre-interaction chirp ($L_0$) is taken into account in the condition $\Gamma_0 \ll 1$. These experimental realizations require the well-designed controllability and measurability of pre-interaction drift ($L_0$), the relative phase ($\phi_0$) for light-electron interaction, and the suppression of decoherence, which definitely challenge the state-of-art techniques of ultrafast electron microscopes [6].

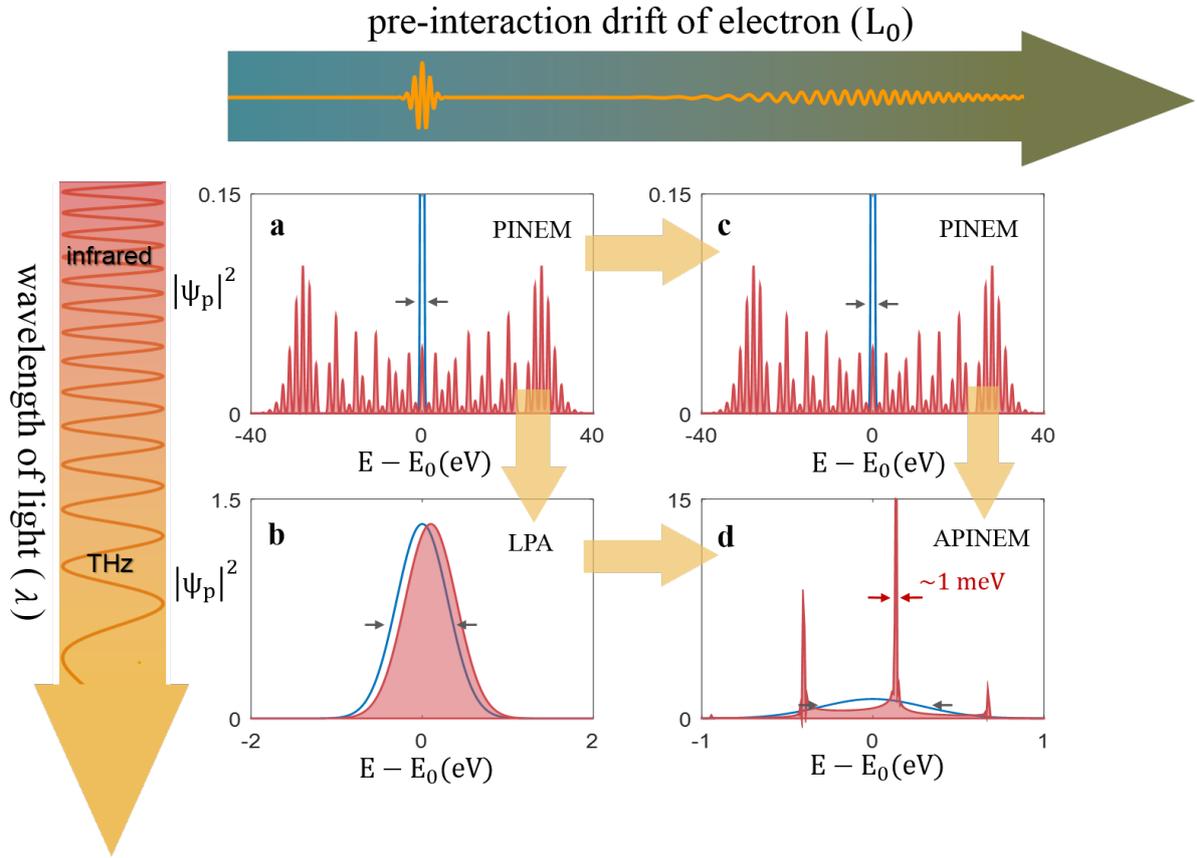

Figure 4| The relationship between the PINEM, LPA and APINEM regimes. (a) The PINEM regime at the condition $\sigma_E < \hbar\omega$, with no pre-interaction chirp $t_0 = 0$ and the strong field strength $|g| = 10$, $\sigma_E = 0.3$ eV. (b) The PINEM regime with the same parameters as in (a) except for the pre-interaction chirp duration $t_0 \neq 0$, where both (a, c) are in the infrared range ($\hbar\omega = 1.55\ eV, \lambda = 800\ nm$). (c) The LPA regime ($\hbar\omega < \sigma_E$, $\sigma_E = 0.3$ eV) with no pre-

interaction drift $t_0 = 0$ in THz regime ($\hbar\omega = 5\ meV, \lambda = 0.25\ mm$); (d) The appearance of an APINEM regime either from acceleration regimes by adding the pre-interaction drift from (b) to (d) (i.e., $L_0 = 0.23\ cm$), or directly from the chirped PINEM by changing the interaction sources from infrared (c) to THz (d). The spectral micro-bunch is in the energy scale of few meV.

**Conclusion** - In short, we analytically revealed three typical quantum regimes in strong field physics: photon-induced near-field electron microscopy (PINEM), linear particle acceleration (LPA) and anomalous PINEM (APINEM). The quantum emergence of laser-driven acceleration in a strong coupling regime and its quantum-to-classical transition are still lacking experimental verifications. Moreover, the optical spectral focusing and periodically spectral bunching of APINEM for improving the energy resolution of EELS detection and spectroscopy remain unexplored, in which these novel phenomena have motivated the present study. Accordingly, we believe our findings can bridge many different fields of fundamental quantum physics, including: strong-field light-matter interaction, laser accelerators, and ultrafast electron microscopy, which are of great interests to these various communities.

**Acknowledgements**

We thanks Ofer Kfir, Michael Kruger for useful discussions. This work was supported by National Natural Science Foundation of China (NSFC) under No.11504038, 'Chongqing Fundamental, Frontier Research Program' (No. cstc2015jcyjA00013), and the Foundation of Education Committees of Chongqing (No. KJ1500411), and is also partially supported by ICORE, Israel Center of Research Excellence program of the Israel Science Foundation (ISF), and by the Crown Photonics Center.

**Supplementary Material:**

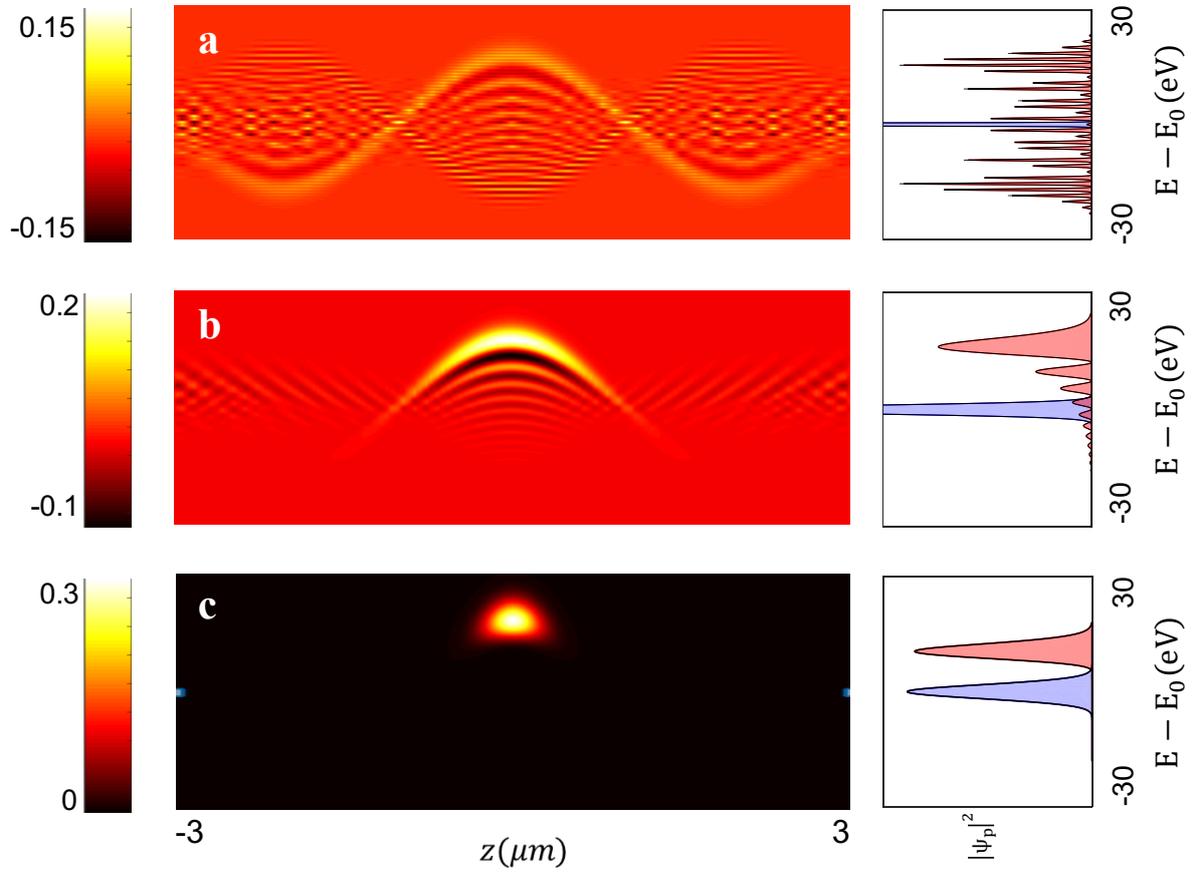

Figure S1| The three typical Wigner function representations in different regimes. (a) PINEM; (b) The transition regime between PINEM and acceleration; (c) The linear acceleration. The simulation parameters are in the ranges $\sigma_E \ll \hbar\omega, \sigma_E \approx \hbar\omega, \sigma_E \gg \hbar\omega$, respectively, where $\hbar\omega = 1.55\ eV, \phi_0 = 0, |g| = 6$.

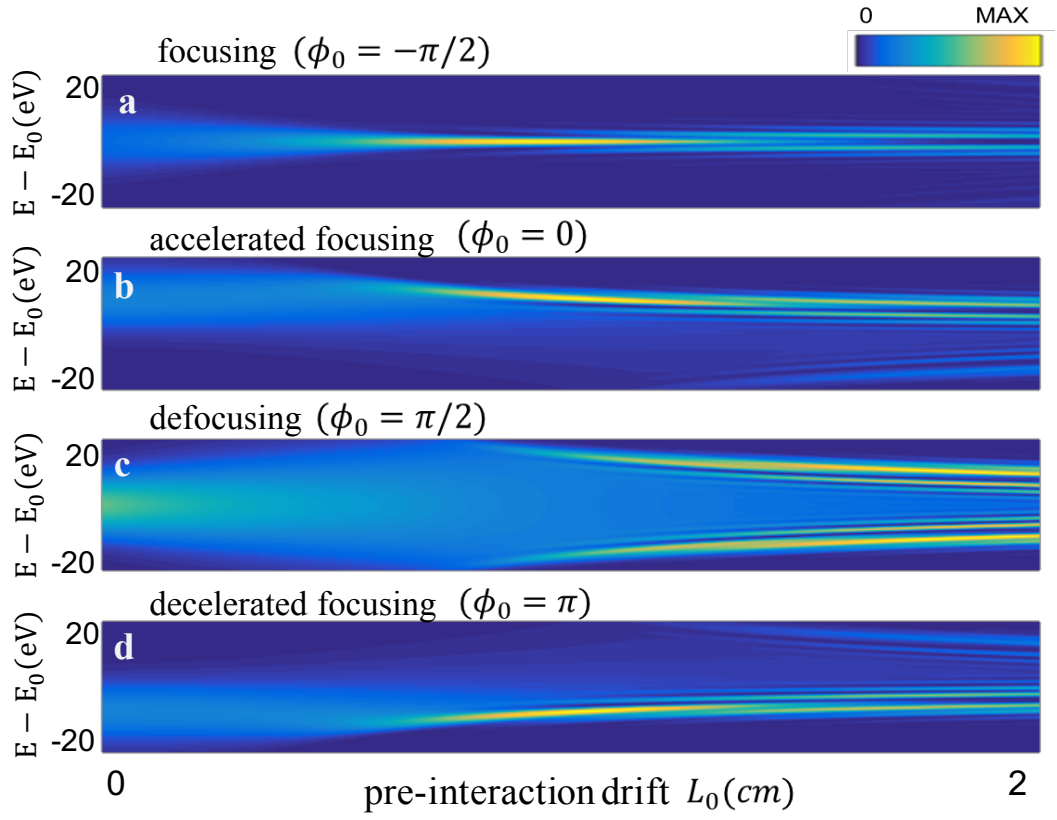

Figure S2| The APINEM regime - spectral focusing, accelerated focusing, defocusing and decelerated focusing cases as a function of pre-interaction drift length $L_0$ with different relative phases $\phi_0 = -\pi/2, 0, \pi/2, \pi$, respectively. The relevant parameters are $\hbar\omega = 1.55\text{eV}$, $\Gamma_0 = 0.1, |g| = 3$.

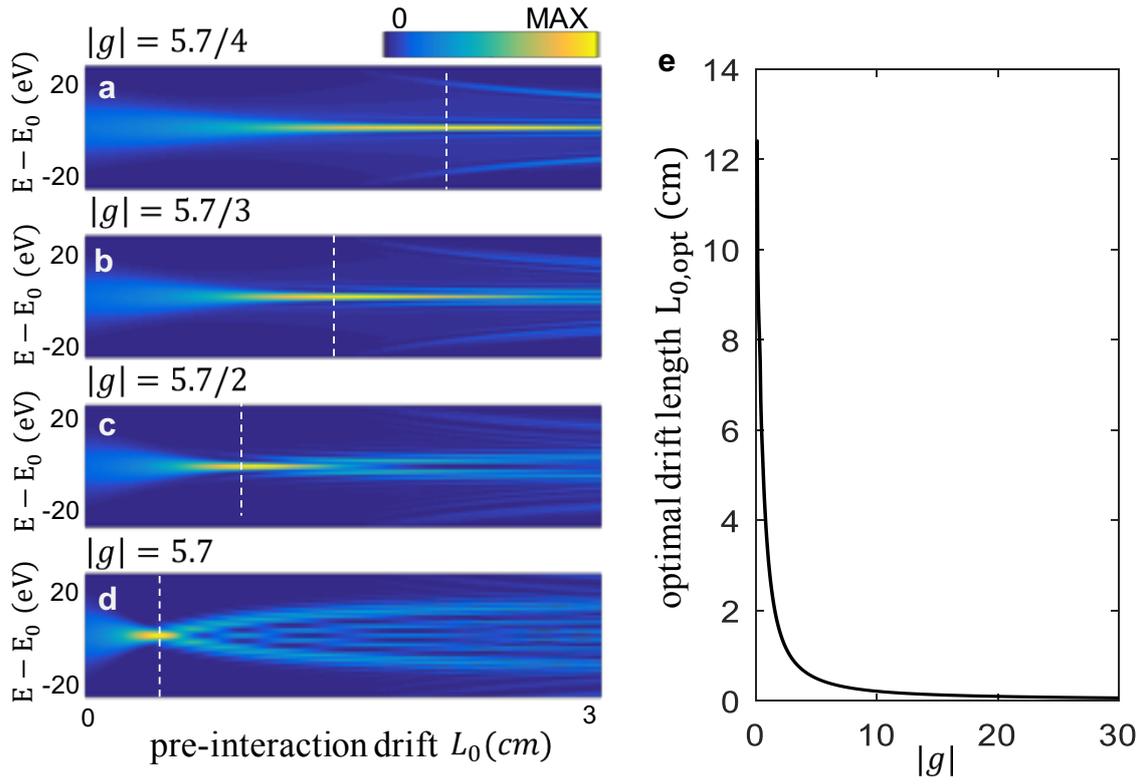

Figure S3| Dependence of the spectral focusing as a function of the pre-interaction drift $L_0$ at different incident field strengths. The optimal focusing length $L_{0,opt}$ exponentially decreases as the incident field strength $|g|$ increases. The relevant parameters are $\hbar\omega = 1.55\text{eV}$, $\Gamma_0 = 0.1$.